\documentclass[10pt,a4paper]{article} 

\usepackage{graphicx}
\usepackage{amsmath,amsthm,bm}

    \DeclareMathOperator\rad{rad}
    \DeclareMathOperator\GF{GF}
    \DeclareMathOperator\GL{GL}
    \DeclareMathOperator\diag{diag}
    \DeclareMathOperator\PG{PG}

\newcommand{\vA}{{\bm A}}
\newcommand{\vX}{{\bm X}}
\newcommand{\vY}{{\bm Y}}

\newcommand{\cI}{{\mathcal I}} 
\newcommand{\cN}{{\mathcal N}}

\newcommand\negativespace{\hspace{0mm}}

\newcommand{\spmatrix}[1]
{\mbox{\scriptsize\setlength\arraycolsep{0.5\arraycolsep}$\begin{pmatrix}#1\end{pmatrix}$}}

\newtheorem{lem}{Lemma}
\newtheorem{thm}{Theorem}

\begin{document}

\title{Vectors, Cyclic Submodules and Projective Spaces Linked with Ternions}

\author{Hans Havlicek \and Metod Saniga}

\maketitle

\begin{abstract}
Given a ring of ternions $R$, i.\,e., a ring isomorphic to that of upper
triangular $2\times 2$ matrices with entries from an arbitrary commutative
field $F$, a complete classification is performed of the vectors from the free
left $R$-module $R^{n+1}$, $n \geq 1$, and of the cyclic submodules generated
by these vectors. The vectors fall into $5 + |F|$ and the submodules into $6$
distinct orbits under the action of the general linear group $\GL_{n+1}(R)$.
\par
Particular attention is paid to {\it free} cyclic submodules generated by
\emph{non}-unimodular vectors, as these are linked with the lines of
$\PG(n,F)$, the $n$-dimensional projective space over $F$. In the finite case,
$F$ = $\GF(q)$, explicit formulas are derived for both the total number of
non-unimodular free cyclic submodules and the number of such submodules passing
through a given vector. These formulas yield a combinatorial approach to the
lines and points of $\PG(n,q)$, $n\geq 2$, in terms of vectors and
non-unimodular free cyclic submodules of $R^{n+1}$.
\end{abstract}

\section{Introduction}
Projective spaces over rings (see \cite{veld} for the standard terminology,
notation and the necessary background information), and projective lines in
particular (see \cite{bh05}), have recently become the subject of considerable
interest due to rather unexpected recognition of their relevance for the field
of quantum physics in general and quantum information theory in particular; we
refer to \cite{hs08}, \cite{pb07}, and the references therein. Being motivated
by these intriguing applications, we have had---in the framework of a broader
international collaboration---a detailed look at the structure of a variety of
finite projective ring lines and planes (see, e.\,g.\ \cite{spkp07}) and came
across some interesting aspects (see, among others, \cite{san07}) which, to the
best of our knowledge, have not yet been the subject of a systematic
mathematical treatment. These aspects mostly relate to the properties of free
cyclic submodules generated by vectors of a free $R$-module of a given unital
ring $R$, and can be summarised into the following open problems: how the
interrelation between different free cyclic submodules over a particular ring
is encoded in the structure of the ideals of the ring; what kind of finite
rings feature ``outliers'', i.\,e., vectors not belonging to any free cyclic
submodule generated by unimodular vectors; what the conditions are for a
non-unimodular vector to generate a free cyclic submodule; and, finally, how
the substructure generated by such non-unimodular free cyclic submodules
relates to the parent ring geometry. These questions lead to projective lattice
geometry in the sense of \cite{bgs}. In order to partially answer some of them,
we have already examined the case of the smallest ring of ternions
\cite{shpp08}---this being, remarkably, the {\it lowest} order ring where one
not only finds ``outliers'', but also free cyclic submodules generated by (some
of) them. In the present paper we extend and generalise the findings of
\cite{shpp08} to an arbitrary ring of ternions, with finite cases handled in
somewhat more detail.

\section{Ternions}

Let $F$ be a (commutative) field. We denote by $R$ the ring of \emph{ternions},
i.\,e., \emph{upper triangular $2\times 2$ matrices} over $F$, with the usual
addition and multiplication for matrices. The ring $R$ is non-commutative, with
$I$ (the $2 \times 2$ identity matrix over $F$) being its multiplicative
identity and 0 (the $2 \times 2$ zero matrix over $F$) the additive
one.\footnote{In what follows, the symbol ``0'' stands, by abuse of notation,
for both the zero-element of the field $F$ and the zero matrix of $R$, the
difference being always fairly obvious from the context.} The ring $R$ has
precisely two (two-sided) ideals other than $0$ and $R$, namely the sets
\begin{equation}\label{eq:left12}
    \cI_1 :=\left\{\left.\begin{pmatrix} 0 & y\\ 0 & z\end{pmatrix}\right| y,z\in
    F\right\}
    \mbox{~~~and~~~}
    \cI_2     :=\left\{\left.\begin{pmatrix} x & y\\ 0 & 0\end{pmatrix}\right| x,y\in
    F\right\}.
\end{equation}
Furthermore, all sets
\begin{equation}\label{eq:right2b:c}
    \cI_1(b:c):=\left\{\left.\begin{pmatrix} 0 & zb\\ 0 & zc\end{pmatrix}\right| z\in
    F\right\}\;\;\mbox{with}\;\;(0,0)\neq (b,c)\in F^2
\end{equation}
are proper right ideals of $R$. In fact, $\cI_1(b:c)$ depends only on the ratio
$b:c$. Note that the ratio $1:0$ is also allowed here. Similarly, all sets
\begin{equation}\label{eq:left1a:b}
    \cI_2(a:b):=\left\{\left.\begin{pmatrix} xa & xb\\ 0 & 0\end{pmatrix}\right| x\in
    F\right\}\;\;\mbox{with}\;\;(0,0)\neq (a,b)\in F^2
\end{equation}
are proper left ideals of $R$. It is an easy exercise to show that there are no
other proper one-sided ideals in $R$ apart from the ones given by (\ref{eq:right2b:c})
and (\ref{eq:left1a:b}). Recall that the \emph{Jacobson radical\/} of $R$
(denoted by $\rad R$) equals the intersection of all maximal left (or right)
ideals. So
\begin{equation}\label{eq:rad}
    \rad R  = \cI_1(1:0)= \cI_2(0:1).
\end{equation}
Note that a ternion is invertible (a unit) if, and only if, its diagonal
entries are non-zero. The set of invertible ternions will be written as $R^*$.
It is a group under multiplication.

\section{Classifying vectors and cyclic submodules}

We consider now the free left $R$-module $R^{n+1}$ for some integer $n\geq 1$.
(The assumption $n\geq 1$ is needed, for example, to guarantee that all six
cases which appear in the proof of Lemma~\ref{lem:vectors} below actually
occur. We refrain from discussing the trivial cases $n=-1$ and $n=0$ throughout
the paper.) We use boldface letters for vectors and matrices with entries from
$R$. Any $s\times t$ matrix $(A_{ij})$ over $R$ can be considered as a $2s
\times 2t$ matrix over $F$ which is partitioned in $2\times 2$ blocks formed by
the upper triangular matrices $A_{ij}$, and vice versa. Multiplication of
matrices over $R$ is equivalent to multiplication of matrices over $F$ under
this one-one correspondence. Thus, in particular, it is easy to check whether a
square matrix over $R$ is invertible by calculating the determinant of the
associated matrix over $F$.
\par
Our first aim is to classify the (row) vectors of the free left module
$R^{n+1}$ up to the natural action of the general linear group $\GL_{n+1}(R)$.
Given a vector $\vX=(X_0,X_1,\ldots,X_n)\in R^{n+1}$ let $\cI_\vX$ denote the
\emph{right\/} ideal of $R$ which is generated by $X_0,X_1,\ldots,X_n$.

\begin{lem}\label{lem:vectors}
 Two vectors $\vX,\vY\in
R^{n+1}$ are in the same $\GL_{n+1}(R)$-orbit if, and only if, the right ideals
$\cI_\vX$ and $\cI_{\vY}$ coincide.
\end{lem}

\begin{proof}
If a vector $\vX\in R^{n+1}$ is multiplied by a matrix $\vA\in\GL_{n+1}(R)$,
then the coordinates of $\vY:=\vX\cdot \vA$ belong to $\cI_\vX$. By virtue of
the inverse matrix $\vA^{-1}$, we see that actually $\cI_\vX=\cI_{\vY}$.

\par
In order to show the converse, we establish that the orbit of any vector
$\vX=(X_0,X_1,\ldots,X_n)$ contains a distinguished vector which depends only
on the right ideal $\cI_\vX$. In our discussion below we make use of two
obvious facts. Firstly, we may permute the coordinates of a vector in an
arbitrary way by multiplying it with a permutation matrix. Secondly, if $\vX$
is a non-zero vector with $X_0\neq 0$, say, then one of the entries of the
ternion $X_0$ is a scalar $w\neq 0$. Multiplying $\vX$ by the invertible matrix
$\vA=\diag(w^{-1}I,I,\ldots,I)$ gives a vector $\vX':=\vX\cdot\vA$ such that
the entry of the ternion $X'_0$ at the same position equals $1\in F$. So,
without loss of generality, we may assume $w=1$ from the very beginning.
\par
\mbox{Case~1:~}$\cI_\vX=0$, so that $\vX=(0,0,\ldots,0)$ is already the
distinguished vector.
\par

\mbox{Case~2:~}$\cI_\vX=\rad R$. Thus $\vX$ has the form $(X_0,X_1,\ldots,X_n)$
with $X_i=\spmatrix{0&y_i\\0&0}$ and $y_0=1$, say. Multiplying $\vX$ by the
invertible matrix
\begin{equation*}
    \begin{pmatrix}I&-y_{1}I&\ldots&-y_{n}I\\
                        0&I&\ldots&0\\
                        \multicolumn{4}{c}{\dotfill}\\
                        0&0&\ldots&I
    \end{pmatrix}
\end{equation*}
gives the distinguished vector $\left(\spmatrix{0&1\\0&0},0,\ldots,0\right)$.
\par

\mbox{Case~3:~}$\cI_\vX=\cI_1(b:1)$ for some $b\in F$. Hence we may assume
$\vX=(X_0,w_1X_0,\ldots,w_nX_0)$ with $X_0=\spmatrix{0&b\\0&1}$ and
$w_1,\ldots,w_n\in F$. Multiplying $\vX$ by the invertible matrix
\begin{equation*}
    \begin{pmatrix}I&-w_{1}I&\ldots&-w_{n}I\\
                        0&I&\ldots&0\\
                        \multicolumn{4}{c}{\dotfill}\\
                        0&0&\ldots&I
    \end{pmatrix}
\end{equation*}
gives the distinguished vector $\left(\spmatrix{0&b\\0&1},0,\ldots,0\right)$.
The scalar $b$ which appears in this vector depends only on the right ideal
$\cI_1(b:1)$.
\par

\mbox{Case~4:~}$\cI_\vX=\cI_1$, whence at least one coordinate of $\vX$ has to
be off the Jacobson radical. So the coordinates of $\vX$ read
$X_i=\spmatrix{0&y_i\\0&z_i}$ with $z_0=1$, say. We introduce the shorthand
\begin{equation*}
    d_i:=\det\begin{pmatrix}y_0&y_i\\z_0 & z_i\end{pmatrix}=y_0z_i-y_i
    \mbox{~~for~~}i=1,2,\ldots,n
\end{equation*}
and proceed in two steps as follows: Define
\begin{equation*}
\setlength\arraycolsep{0.5\arraycolsep}
    \vX\cdot \begin{pmatrix}I&-z_{1}I&\ldots&-z_{n}I\\
                        0&I&\ldots&0\\
                        \multicolumn{4}{c}{\dotfill}\\
                        0&0&\ldots&I
    \end{pmatrix} =
    \left(\!\begin{pmatrix}0&y_0\\0&1\end{pmatrix},
    \begin{pmatrix}0&-d_1\\0&0\end{pmatrix},
    \ldots,
    \begin{pmatrix}0&-d_n\\0&0\end{pmatrix}\!
    \right)=:\vX'.
\end{equation*}
Since $\cI_\vX\neq \cI_1(y_0:1)$, at least one of $d_1,d_2,\ldots,d_n$, say
$d_1$, is unequal to $0$. Now multiplying $\vX'$ by the invertible matrix
\begin{equation*}\renewcommand{\arraystretch}{1.2}
    \begin{pmatrix}I& 0 & 0&\ldots& 0\\
                   y_0d_1^{-1}I&-d_1^{-1}I&-d_2 d_1^{-1}I&\ldots&-d_n d_1^{-1}I\\
                        0&0&I&\ldots&0\\
                        \multicolumn{5}{c}{\dotfill}\\
                        0&0&0&\ldots&I
    \end{pmatrix}
\end{equation*}
gives the distinguished vector $\left(\spmatrix{0&0\\0&1},
\spmatrix{0&1\\0&0},0,\ldots,0\right)$.
\par

\mbox{Case~5:~}$\cI_\vX=\cI_2$. We may assume that
$X_i=\spmatrix{x_i&y_i\\0&0}$ and $x_0=1$, because at least one coordinate of
$\vX$ has to be off the Jacobson radical. Multiplying $\vX$ by the invertible
matrix
\begin{equation*}
    \begin{pmatrix}A_{00}&-X_{1}&\ldots&-X_{n}\\
                        0&I&\ldots&0\\
                        \multicolumn{4}{c}{\dotfill}\\
                        0&0&\ldots&I
    \end{pmatrix},
    \mbox{~~where~~}
    A_{00}:=\begin{pmatrix}1&-y_0\\0&1\end{pmatrix},
\end{equation*}
gives the distinguished vector $\left(\spmatrix{1&0\\0&0},0,0,\ldots,0\right)$.
\par

\mbox{Case~6:~}$\cI_\vX=R$. Assume, first, that none of the entries of $\vX$ is
invertible. Hence we have, for example, $X_0\in \cI_2\setminus \cI_1$ and
$X_1\in \cI_1\setminus \cI_2$. Define
\begin{equation*}
    \vX\cdot\begin{pmatrix}I&0&0&\ldots&0\\
                        I&I&0&\ldots&0\\
                        0&0&I&\ldots&0\\
                        \multicolumn{5}{c}{\dotfill}\\
                        0&0&0&\ldots&I
    \end{pmatrix}=:\vX'.
\end{equation*}
Then $X'_0=X_0+X_1$ is a unit. Thus, we may restrict ourselves to the case when
one of the entries of $\vX$, say $X_0$, is a unit. Now, multiplying $\vX$ by
the invertible matrix
\begin{equation*}
    \begin{pmatrix}X_0^{-1}&-X_0^{-1}X_1&-X_0^{-1}X_2&\ldots&-X_0^{-1}X_n\\
                        0&I&0&\ldots&0\\
                        0&0&I&\ldots&0\\
                        \multicolumn{5}{c}{\dotfill}\\
                        0&0&0&\ldots&I
    \end{pmatrix}
\end{equation*}
gives the distinguished vector $(I,0,\ldots,0)$.
\end{proof}
The previous proof shows that for $n\geq 1$ the vectors of $R^{n+1}$ fall into
$5+|F|$ orbits.
\begin{lem}\label{lem:points}
Under the action of the general linear group $\GL_{n+1}(R)$, $n\geq 1$, the
cyclic submodules of $R^{n+1}$ fall into six orbits with the following
representatives:
\renewcommand\negativespace{\hspace{-1.5mm}}
\begin{eqnarray}\renewcommand\arraystretch{1.2}
 R(0,0,\ldots,0)
    &=&
 \{(0,0,\ldots,0)\}.\label{eq:cs1}
\\
 R\left(\spmatrix{0 &1\\0&0},0,\ldots,0\right)
    &=&
 \left\{\left.\left(\spmatrix{0&y\\0&0},
        0,\ldots,0\right)\right|y\in F\right\}.\label{eq:cs2}
\\
 R\left(\spmatrix{0&0\\0&1},0,\ldots,0\right)
    &=&
 \left\{\left.\left(\spmatrix{0&y\\0&z},
        0,\ldots,0\right)\right|y,z\in F\right\}.\label{eq:cs3}
\\
 R\left(\spmatrix{0&0\\0&1},
        \spmatrix{0&1\\0&0},0,\ldots,0\right)
    &=&
 \left\{\left.\left(\spmatrix{0&y\\0&z},
                    \spmatrix{0&x\\0&0},
                     0,\ldots,0\right)\right|x,y,z\in F\right\}.\label{eq:cs4}
\\
 R\left(\spmatrix{1&0\\0&0},0,\ldots,0\right)
    &=&
 \left\{\left.\left(\spmatrix{x&0\\0&0},
 0,\ldots,0\right)\right|x\in F\right\}.\label{eq:cs5}
\\
 R\left(I,0,\ldots,0\right)
    &=&
 \left\{\left.\left(\spmatrix{x&y\\0&z},
 0,\ldots,0\right)\right|x,y,z\in F\right\}.\label{eq:cs6}
\end{eqnarray}
\end{lem}
\begin{proof}
The assertion follows immediately from the classification of vectors in
Lemma~\ref{lem:vectors}.
\end{proof}
It is worth noting that the cyclic submodule given in (\ref{eq:cs3}) is
generated by any vector $\vX$ with $\cI_\vX=\cI_1(b:1)$ for an
\emph{arbitrary\/} $b\in F$. This illustrates once more that the classification
of cyclic submodules is coarser than the classification of vectors.
\par
In the terminology of \cite[p.~1129]{bgs} the cyclic\footnote{In \cite{bgs}
such submodules are called \emph{$1$-generated\/} rather than ``cyclic''. The
latter term has a different meaning there \cite[p.~1121]{bgs}.} submodules of
$R^{n+1}$ are the \emph{points\/} of the \emph{projective lattice geometry\/}
given by $R^{n+1}$. The only \emph{free points}, i.\,e., free cyclic
submodules, appearing in Lemma~\ref{lem:points} are given in (\ref{eq:cs4}) and
(\ref{eq:cs6}). The point in (\ref{eq:cs6}) is \emph{unimodular}, because there
exists an $R$-linear form $R^{n+1}\to R$ which takes $(I,0,\ldots,0)$ to $I\in
R$. The point in (\ref{eq:cs4}) is not unimodular, since none of its vectors is
mapped to $I\in R$ under an $R$-linear form. We shall not be concerned with the
orbits of the remaining points from Lemma~\ref{lem:points}, since none of them
is free.
\par
The orbit of (\ref{eq:cs6}) under the action of $\GL_{n+1}(R)$ is thus the set
of ``ordinary'' (i.\,e., unimodular free) points. Note that only the elements
of this set are called ``points'' in \cite{dep59}, \cite{dep60}, and
\cite{veld}. See also the section on Barbilian spaces in projective lattice
geometries in \cite[p.~1135--1136]{bgs}. The orbit of (\ref{eq:cs4}) gives rise
to the set of ``extraordinary'' (i.\,e., non-unimodular free) points of the
projective lattice geometry associated with $R^{n+1}$. It is the latter set we
shall consider in the sequel, due to its link with the $n$-dimensional
projective space $\textrm{PG}(n, F)$.

\section{Linking non-unimodular free cyclic submodules with lines of
$\PG(n,F)$}

The free $R$-left module $R^{n+1}$ turns into a $3(n+1)$-dimensional vector
space over $F$ by restricting the ring of scalars from $R$ to $F$. (We tacitly
do not distinguish between $x\in F$ and the ternion $x\cdot I\in R$.) Each
$R$-submodule of $R^{n+1}$ is at the same time an $F$-subspace of this vector
space. In this section we focus our attention to the $R$-submodule $(\rad
R)^{n+1}\subset R^{n+1}$. There exists an obvious $F$-linear bijection between
$(\rad R)^{n+1}\to F^{n+1} $ which is given by
\begin{equation}\label{eq:coo}
    \left(\begin{pmatrix}0&y_0\\0&0\end{pmatrix},
          \begin{pmatrix}0&y_1\\0&0\end{pmatrix},
          \ldots
            \begin{pmatrix}0&y_n\\0&0\end{pmatrix}
    \right)
    \mapsto (y_0,y_1,\ldots,y_n) .
\end{equation}
In the proof of the following result we use this mapping to consider $(\rad
R)^{n+1}$ as an underlying vector space for the projective space $\PG(n, F$).
Note that we \emph{cannot\/} take $\rad R$ (together with addition and
multiplication from $R$) as the field of scalars for this vector space, but we
have to let $F$ play this role.

\begin{thm}\label{thm:1}
Let $R$ be the ring of ternions over a field $F$. The lines of the projective
space $\PG(n, F)$, $n\geq 1$, are precisely the intersections of $(\rad
R)^{n+1}$ with the non-unimodular free cyclic submodules of the module
$R^{n+1}$.
\end{thm}

\begin{proof}
Under the action of $\GL_{n+1}(R)$ on $R^{n+1}$ the set $(\rad R)^{n+1}$ is
invariant. Consequently, $\GL_{n+1}(R)$ acts as an $F$-linear transformation on
$(\rad R)^{n+1}$. Conversely, given an $F$-linear bijection of $(\rad R)^{n+1}$
it will correspond, in terms of the coordinates given by (\ref{eq:coo}), to a
unique matrix $(a_{ij})\in\GL_{n+1}(F)$. This matrix over $F$ determines the
matrix $(a_{ij}I)\in\GL_{n+1}(R)$ which in turn induces the given
transformation on $(\rad R)^{n+1}$. Thus the $F$-linear bijections of $(\rad
R)^{n+1}$ are precisely the $R$-linear bijections of $R^{n+1}$ restricted to
$(\rad R)^{n+1}$. We add in passing that this action of $\GL_{n+1}(R)$ on
$(\rad R)^{n+1}$ is not faithful.
\par
The non-unimodular free cyclic submodule (\ref{eq:cs4}) meets $(\rad R)^{n+1}$
in a two-dimensional $F$-subspace or, said differently, in a line of $\PG(n,
F$). All non-unimodular free cyclic submodules of $R^{n+1}$ (and, likewise, all
lines of $\PG(n, F$)) form an orbit under the action of the group
$\GL_{n+1}(R)$. This proves the assertion.
\end{proof}
The previous result describes only the \emph{lines\/} of $\PG(n, F$) as certain
subsets of $(\rad R)^{n+1}$. However, for $n\geq 2$ this implies that also the
\emph{points\/} of this projective space are known: A subset $p$ of $(\rad
R)^{n+1}$ is a point if, and only if, there exist non-unimodular free cyclic
submodules $R\vX$ and $R\vY$ of $R^{n+1}$ such that
\begin{equation*}
    R\vX\cap (\rad R)^{n+1}\neq R\vY\cap (\rad R)^{n+1},
\end{equation*}
\begin{equation*}
    p=R\vX\cap R\vY\cap (\rad R)^{n+1},\mbox{~~and~~} |p|>1.
\end{equation*}
Thus the vectors of $(\rad R)^{n+1}$ together with the ``traces'' of the
non-unimodular free cyclic submodules of $R^{n+1}$ completely determine the
structure of $\PG(n,F)$ for $n\geq 2$. There is yet another approach to points
which makes use of unimodular free cyclic submodules. It works even for $n\geq
1$: The points of $\PG(n, F$) are precisely the intersections of $(\rad
R)^{n+1}$ with the unimodular free cyclic submodules of the module $R^{n+1}$.
This follows like in the proof of Theorem~\ref{thm:1} from the fact that the
meet of the submodule (\ref{eq:cs6}) with $(\rad R)^{n+1}$ is a point of
$\PG(n, F$), and from the actions of $\GL_{n+1}(R)$ on $R^{n+1}$ and $(\rad
R)^{n+1}$.
\par
Finally, we note that Theorem~\ref{thm:1} also furnishes the proof of the
validity of the conjecture raised in \cite{shpp08} about the connection between
non-unimodular free cyclic submodules of $R^3$, where $R$ is the ring of
ternions over a Galois field $\GF(q)$, and lines of the projective plane
$\PG(2,q)$.

\section{Combinatorics of the finite case}
We assume throughout this section that $F$ is a Galois field $\GF(q)$ with $q$
elements. So $|R|=q^3$, and the number of units in $R$ is given by $|R^*| =
q(q-1)^2$.
\par
Our first aim is to count the numbers $m_i$ of vectors of $R^{n+1}$ fitting
into cases $i=1,2,\ldots, 6$ according to the classification from the previous
section. All of these numbers are non-zero due to our general assumption $n\geq
1$, which is also adopted in this section.
\par
Case~1: The zero-vector is the only vector arising in this case, whence
$m_1=1$.
\par
Case~2: There are $|{\rad R}|^{n+1}=q^{n+1}$ vectors with entries from the
radical of $R$, including the zero-vector. We obtain therefore $m_2=q^{n+1}-1$.
\par
Case~3: First we consider a fixed $b\in F$. Any vector $\vX$ with
$\cI_\vX=\cI_1(b:1)$ has the form
\begin{equation*}
    \left(w_0\begin{pmatrix}0&b\\0&1\end{pmatrix} ,
          w_1\begin{pmatrix}0&b\\0&1\end{pmatrix} ,
          \ldots,
          w_n \begin{pmatrix}0&b\\0&1\end{pmatrix}\right)
\end{equation*}
with $(0,0,\ldots,0)\neq(w_0, w_1,\ldots,w_n)\in F^{n+1}$. As $b$ varies in
$F$, we get $ m_3 = q(q^{n+1}-1)$.
\par
Case~4: There are $|\cI_1|^{n+1}=q^{2(n+1)}$ vectors with entries from the
ideal $\cI_1$, including the vectors from cases 1, 2, and 3. Hence
\begin{equation*}
    m_4 = q^{2(n+1)} - m_3 - m_2 - m_1 = q(q^{n} -1)(q^{n+1}-1).
\end{equation*}
\par
Case~5: We proceed as before and obtain from $|\cI_2|=q^2$ that
\begin{equation*}
    m_5 = q^{2(n+1)} - m_2 -m_1 = q^{n+1}(q^{n+1}-1).
\end{equation*}
\par
Case~6: All remaining vectors fall into this case. We read off from
$|R^{n+1}|=q^{3(n+1)}$ that
\begin{equation*}
    m_6= |R^{n+1}| - m_5-m_4-m_3-m_2-m_1 = q^{n+1}(q^{n+1}-1)^2.
\end{equation*}

We consider now the set $\cN$ of all vectors of $R^{n+1}$ which belong to at
least one non-unimodular free cyclic submodule of $R^{n+1}$. For this set to be
non-empty we must have $n\geq 1$. Under these circumstances the set $\cN$
comprises precisely the $q^{2(n+1)}$ vectors fitting into cases 1--4 according
to our classification.

\begin{figure}[ht!]
\centerline{\includegraphics[width=.65\textwidth]{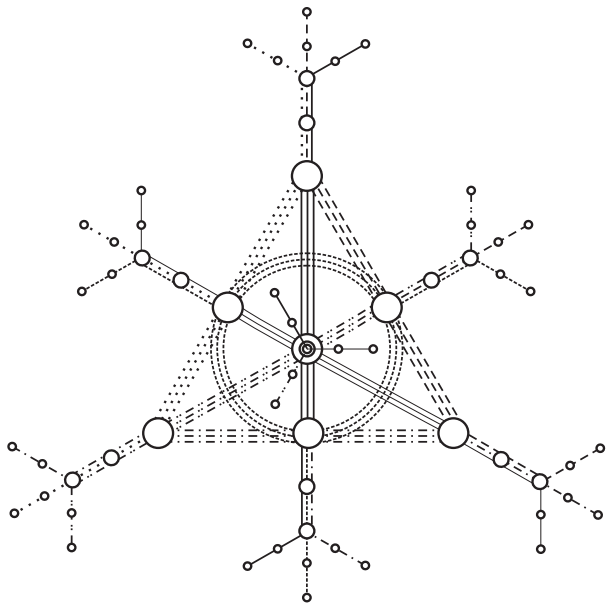}} 

\caption{The ``Fano-Snowflake''---a diagrammatic illustration of the $n=q=2$
case. The $2^6-1 = 63$ vectors of $\cN\setminus\{(0,0,0)\}$ are represented by
circles, whose sizes reflect the number of non-unimodular free cyclic
submodules of $R^{3}$ they are contained in. As the zero vector is not shown,
each submodule of this kind is represented by $2^3-1=7$ circles lying on a
common polygon; three big, two medium-sized, and two small circles
corresponding, respectively, to the vectors from case~2 ($\mu_2=9$), case~3
($\mu_3=3$), and case~4 ($\mu_4=1$). The patterns of the $\mu=21$ polygons were
chosen in such a way to make (the lines of) the Fano plane $\PG(2,2)$ sitting
in the middle of the ``snowflake'' readily discernible. The illustration is
essentially three-dimensional. There is also a single ``vertical branch'' of
the ``snowflake'' which emanates from the the middle of the figure and ramifies
into three smaller branches.}\label{fig:snowflake}
\end{figure}

\begin{thm}\label{thm:2}
Let $R$ be the ring of ternions over $\GF(q)$. There are precisely
\begin{equation}\label{eq:NP}
    \mu:=\frac { ( {q}^{n}-1 )  ( {q}^{n+1} -1 ) }{ ( q-1 )^{2}}
\end{equation}
non-unimodular free cyclic submodules in $R^{n+1}$ for $n\geq 1$. The number of
such submodules containing a vector $\vX\in R^{n+1}$ equals
\begin{eqnarray}
    \label{eq:NP:0}
    \mu_1:=\mu & \mbox{~~{\rm if}~~} &
    \cI_\vX = 0,\\
    \label{eq:NP:rad}
   \mu_2:= \frac { \left( q+1 \right)  \left( {q}^{n}-1 \right) }{q-1} & \mbox{~~{\rm if}~~} &
    \cI_\vX = \rad R,\\
    \label{eq:NP:I_1(b:1)}
   \mu_3:= \frac {{q}^{n}-1}{q-1}& \mbox{~~{\rm if}~~} &
    \cI_\vX = \cI_1(b:1) \mbox{~~{\rm for some}~~}b\in F,\\
   \label{eq:NP:I_1}
   \mu_4:=1 & \mbox{~~{\rm if}~~} &
    \cI_\vX = \cI_1.
\end{eqnarray}
\end{thm}
\begin{proof}
Equation (\ref{eq:NP:I_1}) holds trivially. The number of non-unimodular free
cyclic submodules is $m_4/{|R^*|}$ which yields (\ref{eq:NP}) and
(\ref{eq:NP:0}). In order to establish (\ref{eq:NP:rad}) we count in two ways
the number of pairs $(\vX,R\vY)$, where $R\vY$ is a non-unimodular free cyclic
submodule and the vector $\vX\in R\vY$ is subject to $\cI_\vX=\rad R$.
\par
First we fix $\vY$ and count the number of appropriate vectors $\vX$. Since all
vectors $\vY$ with $\cI_\vY = \cI_1$ are in one orbit of $\GL_{n+1}(R)$, it is
sufficient to consider as $\vY$ the distinguished vector obtained in case~4. By
(\ref{eq:cs4}), the vectors of $R\vY$ have the form
\begin{equation}\label{eq:xyz}
            \left(\begin{pmatrix}0&y\\0&z\end{pmatrix},
            \begin{pmatrix}0&x\\0&0\end{pmatrix},0,\ldots,0\right)
            \mbox{~~with~~} x,y,z\in F.
\end{equation}
Such a vector has the required properties if, and only if, $z=0$ and $(x,y)\neq
(0,0)$. Hence there are $q^2-1$ vectors of this kind. Consequently, as $\vY$
varies also, the number of pairs $(\vX,R\vY)$ is equal to
\begin{equation*}
       (q^2-1)\cdot\mu =  \frac {(q+1) ( {q}^{n}-1 )  ( {q}^{n+1} -1 ) }{ ( q-1 )}.
\end{equation*}
Counting in a different way, we find that this number equals $\mu_2 m_2$,
whence indeed
\begin{equation*}
    \mu_2 = \frac {(q+1) ( {q}^{n}-1 ) }{ ( q-1 )}.
\end{equation*}
\par
The proof of (\ref{eq:NP:I_1(b:1)}) can be accomplished in the same fashion.
The coordinates of a vector given in (\ref{eq:xyz}) generate a right ideal
$\cI_1(b:1)$ for some $b\in F$ if, and only if, $z\neq 0$ and $x=0$. The last
condition is due to the fact that second ternion coordinate of the vector
appearing in (\ref{eq:xyz}) has to be a scalar multiple of the first one. Hence
there are $q(q-1)$ vectors of this kind. This gives
\begin{equation*}
       q(q-1)\cdot\mu =  \frac{ q ( {q}^{n}-1 )  ( {q}^{n+1} -1 ) }{ ( q-1 )} =
       \mu_3 m_3 = \mu_3\cdot q ({q}^{n+1}-1 ),
\end{equation*}
from which the formula for $\mu_3$ is immediate.
\end{proof}
We notice that
\begin{eqnarray*}
  \mu=\mu_1  &=& |{\PG(n-1, q)}|\cdot|{\PG(n, q))}|, \\
  \mu_2      &=& |{\PG(n-1, q)}|\cdot|{\PG(1, q)}|, \\
  \mu_3      &=& |{\PG(n-1, q)}|.
\end{eqnarray*}
Thus for $n\geq 2$ these numbers are distinct and all of them are greater than
$\mu_4=1$. Under these circumstances the four types of vectors in $\cN$ can be
distinguished by the \emph{number of non-unimodular free cyclic submodules in
which they are contained}. Consequently, the set $\cN$ and the family of
non-unimodular free cyclic submodules determine the lines and points of the
projective space $\PG(n,q)$ according to Theorem~\ref{thm:1} and the subsequent
remarks. This provides now the theoretical background for the
\emph{Fano-Snowflake} from \cite{shpp08}, which is depicted in
Figure~\ref{fig:snowflake}, and puts the construction from there in a general
context by allowing an arbitrary dimension $n\geq 2$ and any prime power $q$.

\section{Conclusion}
Given the free left $R$-module $R^{n+1}$, $n \geq 1$, of an arbitrary ring of
ternions $R$, we provide a complete classification of the vectors from
$R^{n+1}$ (Lemma~\ref{lem:vectors}) and the cyclic submodules generated by them
(Lemma~\ref{lem:points}), up to the action of the group $\GL_{n+1}(R)$. There
exist altogether $5 + |F|$ distinct orbits of vectors and six (two \emph{free},
one of them \emph{non}-unimodular) ones of submodules. The non-unimodular free
cyclic submodules are linked with the lines of $\PG(n, F$)
(Theorem~\ref{thm:1}). In the finite case, we count explicitly the total number
of non-unimodular free cyclic submodules as well as the cardinalities of their
subsets passing through a given vector (Theorem~\ref{thm:2}). In light of the
fact that there are only few papers on projective geometries over ternions, we
hope that our findings will stir the interest of mathematicians into a more
systematic treatment of exciting open problems in this particular branch of
ring geometries, and will also prove fruitful for envisaged applications of
projective geometries over ternions in quantum physics.

\subsection*{Acknowledgments}
The work was supported by the VEGA grant agency projects Nos. 6070 and 7012, as
well as by the Action Austria-Slovakia project No. 58s2. We thank Petr Pracna
(Prague) for providing us with the electronic version of the figure.

\par~\par

\noindent\emph{Hans Havlicek\\
Institut f\"{u}r Diskrete Mathematik und Geometrie,
Technische Universit\"{a}t\\
Wiedner Hauptstra{\ss}e 8--10/104,
A-1040 Wien, Austria\\
email: havlicek@geometrie.tuwien.ac.at}

\par~\par

\noindent\emph{Metod Saniga\\
Astronomical Institute, Slovak Academy of Sciences\\
SK-05960 Tatransk\'{a} Lomnica, Slovak Republic\\
email: msaniga@astro.sk}

\end{document}